# Deep level emission of ZnO nanoparticles deposited inside UV opal

S. M. Abrarov[1,*], Sh. U. Yuldashev[1], T. W. Kim[2], Y. H. Kwon[1], and T. W. Kang[1]

[1]*Quantum-functional Semiconductor Research Center, Dongguk University,*

*Seoul 100-715, South Korea*

[2]*Advanced Semiconductor Research Center, Division of Electrical and Computer Engineering,*

*Hanyang University, Seoul 133-791, South Korea*

**Abstract**

The temperature-dependent photoluminescence (PL) spectra of zinc oxide (ZnO) nanocrystals deposited inside the ultraviolet (UV) opal were studied. ZnO was grown in the voids between FCC packed $SiO_2$ spheres using spray pyrolysis under ultrasonic vibration in the solution containing a zinc nitrate precursor. The ZnO nanoparticles inside opal matrix with UV photonic band-gap exhibit suppression of the excitonic emission and enhancement of the deep level emission. Suppression of the excitonic lines is due to the inhibition of spontaneous emission, while enhancement and broadening of the DL emission in the green spectral region is due to Purcell effect. The infiltration of ZnO inside the photonic crystal may be a useful technique to increase its emission efficiency in the selected spectral region.

___________________________

* Corresponding author:

Ph:          +(82)-2-2260-3205
Fax:         +(82)-2-2260-3945

E-mail(s):   absanj@yahoo.co.uk
             abrarov@dongguk.edu

# I. Introduction

Zinc oxide (ZnO) is mostly n-type semiconductor, which has many attractive features. Due to its unique properties in electrical and optical characteristics ZnO is considered a liable material for a variety of applications in the visible and near ultraviolet (UV) spectra. ZnO is a chemically stable direct zone semiconductor with wide band-gap (3.32 eV) and relatively high ecxiton binding energy (60 meV). It is transparent to most of the solar spectrum and therefore ZnO thin films are widely used in industry as a window material for low-cost solar cells, optical wave-guides, light modulators, and optical sensors. The recent publications show that ZnO can be successfully applied as a nanowire Schottky diode [1] and a transparent field effect transistor [2], which can find its application as a switching element in displays. In addition, ZnO is regarded a promising semiconductor for fabrication of the various optoelectronic devices with high light emission efficiency and durability.

Nowadays such advanced techniques as the pulsed laser deposition [3], magnetron sputtering [4], chemical vapour deposition (CVD) [5] and molecular beam epitaxy (MBE) [6] are mostly used to fabricate ZnO. These techniques enable the deposition of single crystal ZnO almost without crystal defects. Due to high crystal quality, ZnO may exhibit only band-edge emission even above room temperature. However all of them are highly expensive and their applications may be cost-effective for the fabrication of single crystal ZnO.

However polycrystalline ZnO may also be very useful for many applications. For instance, polycrystalline ZnO has recently attracted much attention as efficient luminescent phosphor for the low-voltage flat panel displays [7-9]. The simple, inexpensive and safe technique, which can serve for such a purpose is spray pyrolysis (SP).

SP is very technological utilizing just a solvent with precursor such as zinc acetate. It is particularly effective for the deposition of various semiconductors with nanosize particles. Recent publications show that it can be successfully applied for fabrication of the bright color phosphors [10-12]. Furthermore, SP employs a precursor solution whereby the uniform distribution of nanoparticles in the voids of a porous material can be archived by soaking a sample under ultrasonic vibration.

## II. Preliminaries

Various applications of photonic crystals (PhCs), proposed by E. Yablonovich [13] and S. John [14], are aimed to enhance the useful optical characteristics of the opto-electronic devices. Among numerous properties of PhCs, their ability to suppress spontaneous emission band [15] is one of the most remarkable. This effect, the suppression of the radiative recombinations, is observed due to the inhibition of spontaneous emission in the luminescence spectrum, common with photonic band-gap. In fact, the suppression of the emission band leads to the strong spectral redistribution of emission intensity. Therefore the application of PhC may be very effective to modify a spectrum of the spontaneous emission.

Artificial opal is one of the kinds of PhCs, which fabrication is relatively simple and cheap. It is packed into FCC matrix by means of self-sedimentation in a fluid containing sub-micron size spheres (host material). The main disadvantage in fabrication of artificial opal is the long time of self-sedimentation process. Usually for fabrication of the high quality tree-dimensional (3D) matrix, the self-sedimentation process can continue up to several months and sometimes even more than a year.

In the present study the host material of opal matrix is silicon dioxide (SiO$_2$), while the infiltrated luminescent material is ZnO. The photonic band-gap of the bare opal depends on the average diameter of the host material and its index of refraction. Infiltrated luminescent nanoparticles alter the photonic band-gap position because of change in effective index of refraction. It should be noted that the color of opal determines a forbidden spectral zone since the light matching the photonic band-gap is bounced back from its surface.

Unlike other PhCs, the artificial opal can be fabricated to cover a large area over a plane or slightly dipped surface. Moreover, it is quite possible to achieve a filling factor above 90% inside inverted opal, which can be obtained by removing a host material in an acid after initial infiltration. The feasibility to cover a large area and to achieve the high filling factor signifies that the infiltration of ZnO into 3D opal matrix may be very promising for fabrication of bright phosphors for full-color displays.

Typically the PL spectra of ZnO consist of excitonic emission in the near UV region and deep level (DL) emission mostly in the green and partly in the yellow and red spectral regions. The DL emission in ZnO strongly manifests in the polycrystalline structure due to high density of native defects such as oxygen vacancies and Zn interstitials [9, 16, 17].

The PL spectra exhibiting a narrow excitonic emission on ZnO thin (90 nm) film deposited on the surface of artificial opal by electron-beam sputtering was reported by A. N. Gruzintsev et al [18, 19]. Thin ZnO film was sputtered on the surface of artificial opal representing a two-dimensional (2D) periodic array of quantum dots distributed between 279 nm SiO$_2$ spheres (red opal). The narrowing of the excitonic emission originated from ZnO quantum dots was observed due to the quantum confinement effect.

The novel approach towards fabrication a 3D matrix containing FCC packed colloidal spheres of ZnO was developed by E. W. Seelig et al [20]. It was demonstrated that at least a quasi-3D matrix, i.e. few layers of the self-assembled ZnO colloidal spheres, is feasible to achieve by natural sedimentation in the fluid suspension. The main advantage of such a configuration is a very high filing factor of ZnO, which can be increased up to 74%.

The precursor solutions can be used to fill the voids of artificial opal with various nanoparticles such as ZnS, ZnO, fluorescent oxides and chlorides [21]. For instance, the complete 3D periodic array of ZnO can be prepared by chemical deposition from aqueous solution containing a zinc nitrate or zinc acetate precursor followed by thermal annealing [22-28]. The first reports dedicated to ZnO nanoparticles grown inside the volume of the artificial opal showed that the complete filling of the 3D matrix can be used as an efficient technique to modify the PL emission of ZnO [22, 23]. The green porous opal has been chosen as a 3D matrix, since its photonic band-gap overlaps the DLs in the electronic band-gap. As a result of the inhibition of spontaneous emission, the ZnO nanoparticles grown in the voids between 220-240 nm $SiO_2$ spheres exhibit the dominant UV-blue emission, since DL emission is suppressed in the green region.

In earlier publications we assumed that in PL spectra of ZnO nanoparticles deposited inside the artificial opal with ultraviolet (UV) photonic band-gap could be used to observe the reverse effect, i.e. the suppression of the UV-blue emission and enhancement of the green emission. For the photonic band-gap occupying near UV spectrum, the corresponding average diameter of $SiO_2$ spheres should be around ~ 160 nm [23, 26]. Due to optical match between the UV light and photonic band-gap of opal matrix, the radiative recombination rate originating through excitons might be decreased as a result of the inhibition of spontaneous emission.

This paper reports the temperature-dependent PL spectra of the ZnO deposited inside UV opal by using SP under ultrasonic vibration. The PL spectroscopy shows several conclusive factors supporting an assumption that the enhancement of the DL emission is due to the influence of PhC. Particularly, the excitonic emission from ZnO embedded UV opal is suppressed, while its DL emission is strongly enhanced and broadened. Furthermore, the band edge emission from ZnO powder rapidly increases, while the emission from ZnO embedded UV opal does not exhibit excitonic lines with decreasing temperature.

## III. Experimental

*3.1 Synthesis and self-sedimentation of $SiO_2$ spheres*

The sub-micron size spheres of silicon dioxide were synthesized in Stöber-Fink-Bohn process [29] through the hydrolysis of tetraethyl orthosilicate (TEOS) in the ethanol solution mixed with ammonium hydroxide and water. The average diameter of the $SiO_2$ spheres can be controlled varying the temperature of the TEOS hydrolysis and the concentrations of ammonium hydroxide and water. An increase of the ammonium hydroxide results to the larger sizes of the spherical particles. On the contrary, the increase of the water concentration results to the smaller sizes of the spherical particles. Typically, the hydrolysis of TEOS in ethanol solution with ammonium hydroxide and water enables the synthesis of the spherical particles with diameters ranging between 0.02 – 2.0 μm.

The artificial opal was assembled as a FCC packed 3D matrix by natural sedimentation of $SiO_2$ spheres in water suspension. After sedimentation, the water was slowly evaporated at the temperature 60-80 °C. Self-assembled FCC packed matrix is mechanically fragile, therefore it was sintered for several hours at temperature 1000 °C.

This technique enables the fabrication the $SiO_2$ spheres with small, less than 10%, deviation from their average size. The more complete description of the synthesis of the $SiO_2$ spheres in Stöber-Fink-Bohn process and self-assembled 3D matrix formation by natural sedimentation in water can be found elsewhere [30, 31].

*3.2 Spray pyrolysis technique and ZnO formation*

SP is a technique for chemical deposition of various films, based on a spraying of precursor molecules on a heated surface. The process of pyrolysis is the chemical decomposition occurring under high temperature. Particularly ZnO nanoparticles are formed on the surfaces of $SiO_2$ spheres as a result of the chemical decomposition of precursor molecules.

Generally, the deposition of ZnO by means of precursor solution and thermal annealing is also referred as sol-gel [26] and chemical deposition [24, 25, 27]. However, sol-gel may not be precise and complete according to its definition to describe the process of thermal decomposition employing the precursor solution technique. Another name, the chemical deposition, is very general, since it can be referred to any kind of chemical depositions. SP exactly and completely describes the process of formation of ZnO nanoparticles taking place as a result of spraying of precursor molecules and the process of pyrolysis (thermal decomposition).

The complete chemical name of the precursor is zinc nitrate hexahydrate, $Zn(NO_3)_2 \cdot 6H_2O$, which looks like white or slightly greenish powder. ZnO was deposited using 0.1 M deionized water solution containing the precursor of 99.999% purity. The chemical synthesis of ZnO is described by the following schematics:

$$Zn(NO_3)_2 \bullet 6H_2O \rightarrow \begin{pmatrix} Zn(OH)_2 \\ Zn_2O(NO_3)_2 \\ Zn(OH)(NO_3) \end{pmatrix} \rightarrow ZnO + yN_wO_x + zO_2$$

In water solution, zinc nitrate hexahydrate is decomposed to zinc hydrogenous and nitrogenous compounds such as zinc dihydroxide, dizinc oxide nitrate, and zinc hydroxide nitrate, respectively, as shown in the parentheses of the schematics. The proportion between these zinc hydrogenous and nitrogenous compounds depends on temperature of the solution. Under thermal annealing, the zinc hydrogenous and nitrogenous compounds are further decomposed to ZnO, nitrogen oxides, and oxygen molecules. The more detailed description of the chemical synthesis of zinc oxide from aqueous solution containing zinc nitrate precursor can be found elsewhere [17, 32].

To infiltrate a sample, the 1 mm-thick opal slab was initially soaked in aqueous solution with zinc nitrate precursor for 48 h. To remove air from the voids and to improve the fluid flow, the opal slab was soaked under ultrasonic vibration. The long ultrasonic vibration is not desirable since it may destroy the ZnO nanoparticles and damage the sample. Therefore the sonification process was performed only for a limited time, 3 h in each infiltration cycle.

After soaking in the precursor solution, the opal slab was dried in the drying oven at temperature 50 $^{o}$C for 3h and annealed at 500 $^{o}$C for 1 h. The artificial opal is a porous material, therefore the solution is able to penetrated effectively inside the volume resulting in a uniform distribution of ZnO nanocrystals between the $SiO_2$ spheres. It should be noted that the DL emission intensity of ZnO strongly depends on annealing as well as drying temperatures. The filling factor of ZnO grown inside the artificial opal can be controlled by concentration of the precursor and by number of infiltration cycles. However the high concentration of the precursor in aqueous solution may deteriorate the uniformity of ZnO inside the matrix.

To improve the penetration of the precursor solution inside the sample, the ZnO layers were removed from the opal surfaces by polishing and cleaning after each infiltration cycle. Finally, for the proper PL measurements, ZnO layers were removed in diluted HF water solution since their contribution may be strong enough to screen the effect of the photonic band-gap on the PL spectra.

As a reference sample, a ZnO powder was also deposited on the crystalline quartz substrate (α-quartz) under similar conditions.

*3.3 Measurements*

PL and transmittance spectra were obtained using a SPEX spectrometer equipped with 0.75 m grating monochromator. PL measurements were performed under 10 mW helium-cadmium laser excitation source operating at wavelength 325 nm. A xenon-mercury lamp was employed as a broadband light source. A photo-multiplier tube Hamamatsu R943-02 was used as a photo-detector. The filling fraction of ZnO was determined gravimetrically, i.e. by weighing a porous UV opal before and after infiltration on Electronic Analytical Balance ER-182A. The morphology of the ZnO embedded UV opal and nanoparticles were investigated using a scanning electron microscope (SEM) XL-30 PHILIPS. The temperature in the chamber was controlled between 10 and 300 K using the CRYODINE He displex system.

## IV. Results and discussion

Figure 1 shows the general morphology SEM image of the artificial opal. The 3D opal matrix consists of the $SiO_2$ spheres packed into FCC structure. The diameters of the spheres vary between 160 – 175 nm. Inset in Figure 1 shows the SEM image of the typical ZnO nanoparicles deposited by spray pyrolysis method. The structure of the ZnO is polycrystalline, consisting of nanoparticles with sizes ranging between 5 – 20 nm. The average size of the ZnO nanoparticles obtained by means of SP can be less than 10 nm [33].

Figure 2 shows the transmittance characteristics of the UV opal before and after infiltration with ZnO nanoparticles. For the bare UV opal, the influence of the photonic crystal on transmittance manifests as a dip with minimum at 396 nm. However, the effect of the photonic band-gap is not evident for the ZnO embedded UV opal since its transmittance just tends to zero at the wavelengths below 390-400 nm. The absence of the dip in transmittance is due to the fact that the ZnO nanoparticles strongly absorb the incident light. As a result of the band-to-band absorption, the light propagating through the sample containing ZnO nanoparticles rapidly decreases screening the effect of photonic band-gap in transmittance spectrum.

However, the influence of the photonic crystal can be seen detecting the light reflected from the surface of the ZnO infiltrated sample. Inset in Figure 2 shows the reflectance of ZnO embedded UV opal. The light reflected from the artificial opal travels the distance of few $SiO_2$ spherical layers before it bounces back due to photonic band-gap. Therefore, the round trip of the light is not long enough to be absorbed by ZnO nanoparticles. The maximum of the reflectance located at 408 nm is red-shifted with respect to minimum of transmittance of the bare UV opal. The red-shift indicates the presence of ZnO nanoparticles in the voids of opal matrix.

The filling factor can be estimated employing the following equation [34]

$$\lambda = 2D \times 0.816 \left( \sqrt{\varepsilon_{Si_2O}}\, 0.74 + \sqrt{\varepsilon_{ZnO}}\, f + \sqrt{\varepsilon_{air}}\, (0.26 - f) \right) \qquad (1)$$

where $\varepsilon_{SiO_2}$, $\varepsilon_{ZnO}$, $\varepsilon_{air}$ are dielectric constants for $SiO_2$, ZnO and air, respectively, $f$ is a filling factor for ZnO.

Using the relation between dielectric constant and index of refraction for nonmagnetic medium $n = \sqrt{\varepsilon}$, Eq. (1) can be expressed in form

$$\lambda = 2D \times 0.816 \left( n_{SiO_2}\, 0.74 + n_{ZnO}\, f + n_{air}\, (0.26 - f) \right) \qquad (2)$$

where $n_{SiO_2}$, $n_{ZnO}$, and $n_{air}$ are indices of refraction for silicon dioxide, zinc oxide, and air, respectively. The sum in parentheses of Eq. (2) represents the effective index of refraction $n_{eff}$ according to effective medium approximation [35].

For the bare opal, the filling factor is $f = 0$. Therefore a red-shift of photonic band-gap arising as a result of infiltration of ZnO nanoparticles can be found using Eq. (2) as follows

$$\Delta\lambda = \lambda_2 - \lambda_1 = 2Df \times 0.816(n_{ZnO} - n_{air}) \qquad (3)$$

where $\lambda_1$, $\lambda_2$ are wavelengths corresponding to the minimum of transmittance (or maximum of reflectance) characteristics. Taking into consideration that the index of refraction for air is close to unity, the Eq. (3) can be rewritten as

$$f = \frac{\Delta\lambda}{2D \times 0.816(n_{ZnO} - 1)} \qquad (4)$$

The red-shift of photonic band-gap in Eq. (3) can be determined experimentally from the transmittance and reflectance characteristics shown in the Figure 2

$$\Delta\lambda = 408\,nm - 396\,nm = 12\,nm \qquad (5)$$

The indices of refraction, $n_{SiO_2}(\lambda)$ and $n_{ZnO}(\lambda)$, for SiO$_2$ and ZnO are both wavelength-dependent. This dependence is particularly evident for ZnO due to resonance near energy band-gap [36]. In the most of the visible range the index of refraction $n_{ZnO}(\lambda)$ is close to 2. However $n_{ZnO}(\lambda)$ rapidly increases approaching to the blue-UV range. Taking roughly $\lambda_{UV} = 400$ nm, substituting the values of red-shift (5) and corresponding index of refraction $n_{ZnO}(\lambda_{UV}) = 2.24$ into Eq. (4), the approximate filling factor was found to be ~ 2.8%. This magnitude matches reasonably well the filling factor determined gravimetrically (~ 2.3%).

Figure 3 shows the temperature-dependent PL spectra of the ZnO powder. The band edge emission increases with decreasing temperature. The PL spectra exhibit the strong and broad emission peak in the green region near 2.38 eV (520 nm). This emission peak is commonly observed in the spectra of the ZnO bulks, films, and powders and attributed to native defects.

At 300 K, the intensities of the excitonic peak at 3.28 eV and DL peak at 2.38 eV are approximately equal. However at 10 K, the intensity of the excitonic emission peak at 3.37 eV is more than three times higher than the intensity of the DL emission peak. Initially, the increase of the band edge emission is relatively rapid with decreasing temperature. However, the band edge emission increases insignificantly below 80 K.

At 300 K the emission peak located at 3.28 eV is attributed to free exciton ($X_A$). At low temperatures, three distinct emission lines are observed at 3.37, 3.315, and 3.232 eV. The emission line located at 3.37 is attributed to free exciton, while the emission lines located at 3.315 and 3.232 eV are attributed to donor-acceptor pairs [37]. The positions of the excitonic and donor-acceptor pair emission lines for the ZnO powder are those for the bulk ZnO. The blue-shift due to the quantum confinement is not observed since the ZnO nanoparticles are not physically isolated, i.e. they are in a physical contact with each other.

Figure 4 shows the temperature-dependent spectra of ZnO embedded UV opal. The DL emission with maximum near 2.3 eV (540 nm) is strong and very broad. Apart from the green spectrum, the DL emission includes the complete yellow and part of the red spectra. The relatively strong red emission peak is located near 1.68 eV (740 nm). This emission peak is associated with recombination centers in the crystal lattice of ZnO also originating through native defects [38].

Comparing Figures 4 with 2, one can see that the DL emission peak of ZnO embedded UV opal is consistent with absorption minimum in the green region. Therefore the minimum observed near 540 nm in the transmittance spectrum of ZnO embedded UV opal might be related to the absorption of the light through native defects.

Inset in Figure 4 shows the PL spectra of the reference sample and ZnO embedded UV opal. In contrast to the PL spectra of the ZnO powder, the band edge emission of ZnO embedded UV opal remains negligible even at low temperatures, while its DL emission is significantly enhanced and broadened. The magnitudes of FWHM associated with native defects are 102 and 178 nm for the spectra of ZnO powder and ZnO embedded UV opal, respectively. The comparison between the PL spectra of the ZnO embedded UV opal measured at 10 and 300 K shows a slight blue-shift and narrowing of the DL emission with decreasing temperature. In the red region, the emission peak located around 740 nm (1.68 eV) manifests significantly from ZnO embedded UV opal even above room temperature, while the nearly visible red emission peak from ZnO powder appears only at low temperatures.

Presumably, the enhancement and broadening of the DL emission is due to Purcell effect [39, 40]. The photonic band-gap of the UV opal occupies the near UV region overlapping the excitonic emission spectrum. As a result, the band edge emission is suppressed due to the inhibition of spontaneous emission. Since the emission in the near UV region is forbidden, the PL intensity is redistributed and released through allowed radiative recombinations in the spectra outside the photonic band-gap [15] ultimately leading to the enhancement and broadening of the PL intensity in the green region.

It should be noted that the feasibility of deposition of $Mg_xZn_{1-x}O$ nanoparticles employing the similar solution technique has been shown [41]. Since the concentration of Mg can be controlled, the infiltration of the ZnO and its compounds inside the UV artificial opal by means of the precursor solution might be promising for the fabrication of the efficient luminescent phosphors with DL emission extended to the blue region direction.

## V. Conclusion

SP under ultrasonic vibration was used for growth of ZnO nanocrystals in the volume of the UV opal. To fill the voids between 160-175 nm $SiO_2$ spheres with ZnO, the opal matrix was soaked in aqueous solution containing a zinc nitrate precursor and thermally annealed.

The enhancement and broadening of the PL intensity in the green region might be attributed to Purcell effect. Since the band edge emission is forbidden, the emission energy is spectrally redistributed leading to the broadened and enhanced DL emission through native defects like oxygen vacancies and Zn interstitials.

The feasibility of artificial opal to cover the large area over a plane surface indicates that its infiltration with ZnO nanoparticles may be promising for application in full-color displays. Presumably, the $Mg_xZn_{1-x}O$ nanoparticles grown in the voids of UV opal, employing a similar solution technique, might exhibit the DL emission extended towards blue spectrum. Infiltration of the luminescent ZnO inside the artificial opal can be used to enhance its emission efficiency in the selected spectral region.

## Acknowledgements

This work is supported by the Korea Science and Engineering Foundation through the Quantum-functional Semiconductor Research Center, and by the research program and fund of Dongguk University, 2005.

**Figure Captions**

**Fig. 1.** General morphology of the ZnO embedded UV opal. Inset shows SEM image of the ZnO nanoparticles.

**Fig. 2.** Transmittance characteristics of the (a) bare UV opal and (b) ZnO embedded UV opal. Inset shows the reflectance of the ZnO embedded opal.

**Fig. 3.** Temperature-dependent PL spectra of ZnO powder.

**Fig. 4.** Temperature-dependent PL spectra of ZnO embedded UV opal. Inset shows the PL spectra of the (a) ZnO powder and (b) ZnO embedded UV opal.

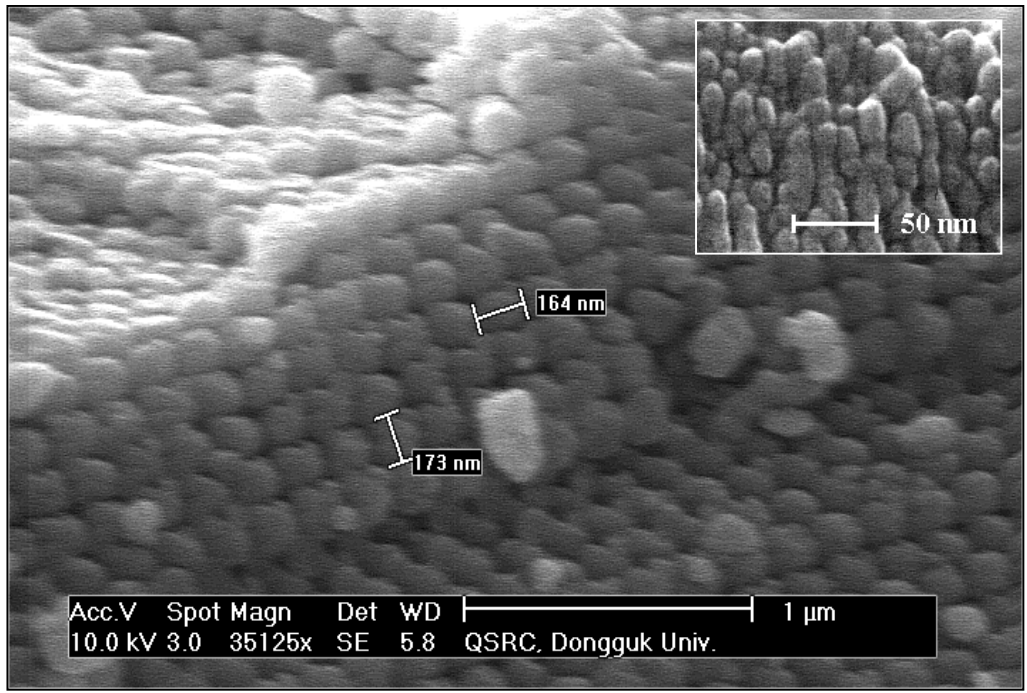

**Fig. 1.** S. M. Abrarov et al.

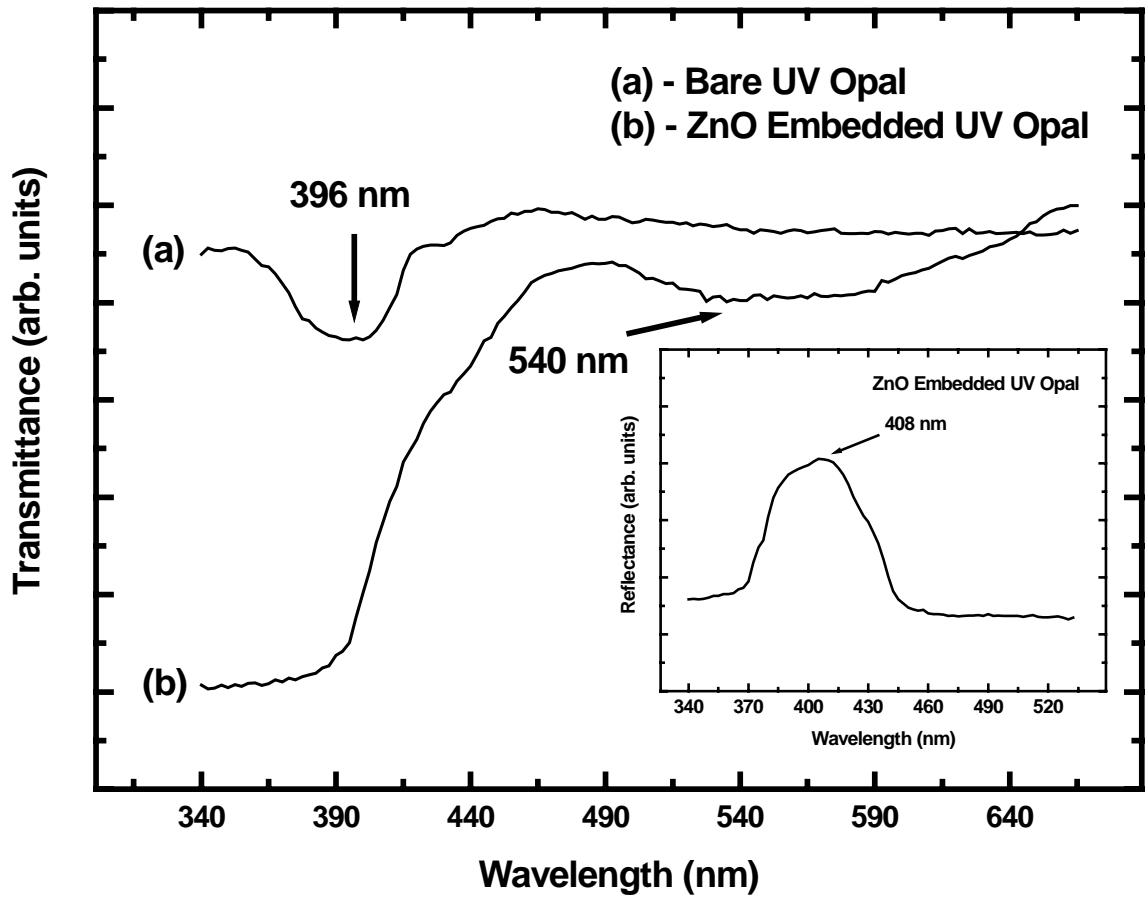

**Fig. 2.** S. M. Abrarov et al.

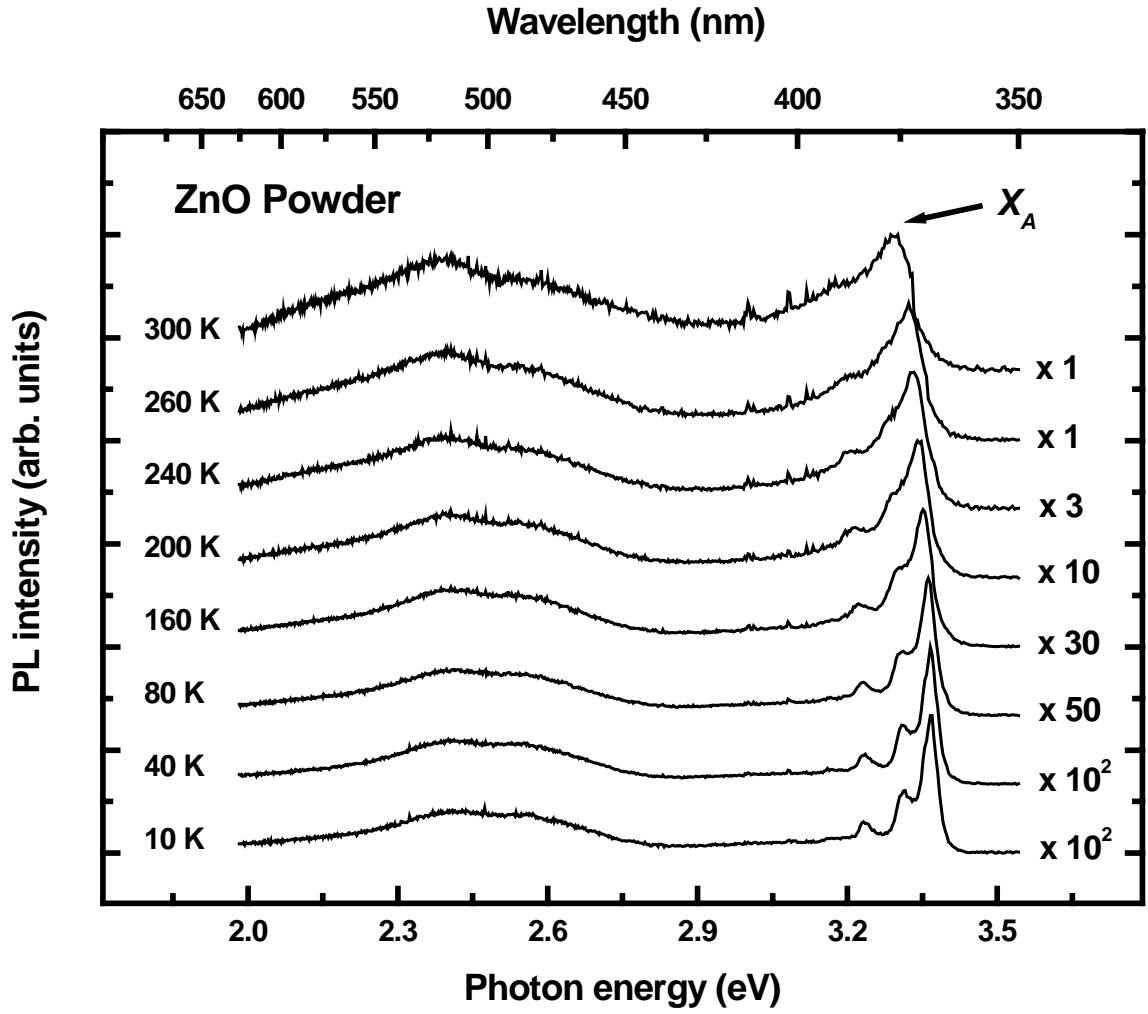

**Fig. 3.** S. M. Abrarov et al.

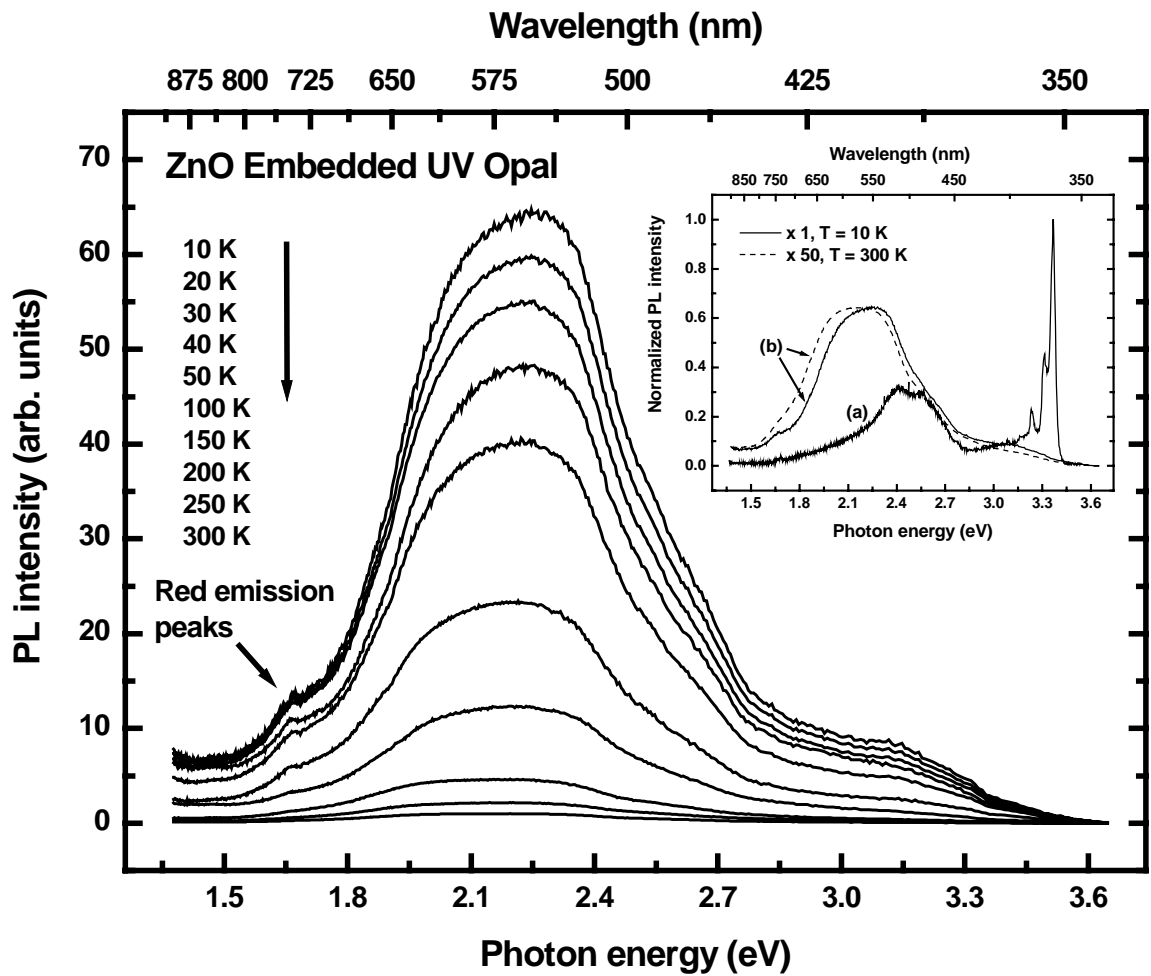

**Fig. 4.** S. M. Abrarov et al.